\documentclass[aps,prl,twocolumn,showpacs,superscriptaddress]{revtex4}
\usepackage{graphicx}

\begin{document}

\title{Spin gap and magnetic resonance in superconducting BaFe$_{1.9}$Ni$%
_{0.1}$As$_{2}$}
\author{Shiliang Li}
\affiliation{Department of Physics and Astronomy, The University of Tennessee, Knoxville,
Tennessee 37996-1200, USA }
\author{Ying Chen}
\affiliation{NIST Center for Neutron Research, National Institute of Standards and
Technology, Gaithersburg, MD 20899, USA }
\author{Sung Chang}
\affiliation{NIST Center for Neutron Research, National Institute of Standards and
Technology, Gaithersburg, MD 20899, USA }
\author{Jeffrey W. Lynn}
\affiliation{NIST Center for Neutron Research, National Institute of Standards and
Technology, Gaithersburg, MD 20899, USA }
\author{Linjun Li}
\affiliation{Department of Physics, Zhejiang University, Hangzhou 310027, China }
\author{Yongkang Luo}
\affiliation{Department of Physics, Zhejiang University, Hangzhou 310027, China }
\author{Guanghan Cao}
\affiliation{Department of Physics, Zhejiang University, Hangzhou 310027, China }
\author{Zhu'an Xu}
\affiliation{Department of Physics, Zhejiang University, Hangzhou 310027, China }
\author{Pengcheng Dai}
\email{daip@ornl.gov}
\affiliation{Department of Physics and Astronomy, The University of Tennessee, Knoxville,
Tennessee 37996-1200, USA }
\affiliation{Neutron Scattering Science Division, Oak Ridge National Laboratory, Oak
Ridge, Tennessee 37831-6393, USA}

\begin{abstract}
We use neutron spectroscopy to determine the nature of the magnetic
excitations in superconducting BaFe$_{1.9}$Ni$_{0.1}$As$_{2}$ ($T_{c}=20$
K). Above $T_{c}$ the excitations are gapless and centered at the
commensurate antiferromagnetic wave vector of the parent compound, while the
intensity exhibits a sinusoidal modulation along the $c$-axis. As the
superconducting state is entered a spin gap gradually opens, whose magnitude
tracks the $T$-dependence of the superconducting gap observed by angle
resolved photoemission. Both the spin gap and
magnetic resonance energies are temperature \textit{and} wave vector
dependent, but their ratio is the same within uncertainties. These results
suggest that the spin resonance is a singlet-triplet excitation related to
electron pairing and superconductivity.
\end{abstract}

\maketitle




The magnetic scattering in the high-transition-temperature (high-$T_{c}$)
copper oxide superconductors is characterized by strong spin correlations in
the vicinity of the antiferromagnetic (AF) wave vector of the magnetically
ordered parent materials, and a spin `resonant' magnetic excitation whose
energy scales with $T_{c}$ and whose intensity develops like the
superconducting order parameter \cite{Mignod,mook,dai,stock}. Like the
cuprates, the Fe-based superconductors \cite{xu,norman,kivelson} are derived from
electron \cite{kamihara,leithe,sefat,ljli} or hole \cite{rotter} doping of
their AF long-ranged ordered parent compounds \cite{cruz,mcguire,jzhao1,qhuang,jzhao2,goldman} and spin fluctuations have been
postulated as the possible glue for mediating the electron pairing for
superconductivity \cite{mazin,chubukov,stanev}. Indeed, the very recent
observation of the same type of magnetic resonant excitation in the
iron-based superconductors \cite{christianson,lumsden,chi} inexorably links
these two high-$T_{c}$ superconductor families together, and strongly
suggests that the pairing mechanism has a common origin that is intimately
tied to the magnetic properties.

An essential step in elucidating the role of magnetism in the
superconductivity of these materials is then an in-depth determination of
the energy ($E=\hbar \omega $) and wave vector ($Q$) dependence of the low
energy magnetic scattering as the superconducting state is formed \cite{mazin,chubukov,stanev}. If electrons in the Fe-based superconductors indeed
form pairs of spin-singlets below $T_{c}$ as in conventional superconductors 
\cite{bcs} and high-$T_{c}$ copper oxides, there can be an energy associated
with exciting the spin-singlet into the high-energy spin-triplet state,
without unbinding the electron pairs. In this picture, the Cooper pairs
should exhibit a wave-vector-independent spin gap with a $T$ dependence that
gradually opens below $T_{c}$, much like the temperature dependence of the
isotropic superconducting gap function observed by angle resolved
photoemission spectroscopy (ARPES) experiments \cite{ding,terashima}. We
have used inelastic neutron scattering to probe the wave vector and energy
dependence of the low energy magnetic excitation spectrum $S(Q,\omega )$. We
find that the spin-gap does open gradually below $T_{c}$, but the gap energy
is dispersive rather than wave-vector independent and tracks the dispersion
of the resonant mode that has been observed \cite{chi}. These results
suggest that the resonant mode is indeed the spin-singlet to spin-triplet
excitation.

We chose single crystals of superconducting BaFe$_{1.9}$Ni$_{0.1}$As$_{2}$
(with $T_{c}=20$ K) because these samples have excellent superconducting
properties \cite{ljli}. In the absence of Ni-doping, BaFe$_{2}$As$_{2}$ is a
nonsuperconducting metal that orders antiferromagnetically with a spin
structure shown in Fig. 1a \cite{qhuang}. Because of the unit cell doubling
along the orthorhombic $a$-axis and $c$-axis spin arrangement, magnetic
Bragg reflections occur at wave vectors $Q=(1,0,1)$ and $(1,0,3)$ type
positions and are absent at $Q=(1,0,0)$ and $(1,0,2)$ \cite{qhuang,jzhao2,goldman}. Previous neutron scattering experiments on
hole-doped Ba$_{0.86}$K$_{0.4}$Fe$_{2}$As$_{2}$ powder samples \cite{christianson} and single crystals of BaFe$_{1.84}$Co$_{0.16}$As$_{2}$ ($T_{c}=22$ K) \cite{lumsden} have shown that the effect of superconductivity
is to induce a neutron spin resonance at energies of $\sim $$5k_{B}T$,
remarkably similar to the doping dependence of the resonance in high-$T_{c}$
copper oxides \cite{wilson,ncco} and heavy fermions \cite{metoki,stock1}.
Measurements on single crystals of BaFe$_{1.9}$Ni$_{0.1}$As$_{2}$ ($T_{c}=20$
K) \cite{chi} suggest that the resonance actually exhibits dispersion along
the $c$-axis, and occurs at distinctively different energies at the
three-dimensional (3D) AF ordering wave vector ${Q}=(1,0,1)$ and at $Q=(1,0,0)$. We note that in the parent materials the spin wave dispersions
in the Fe-based superconductors are anisotropic and clearly 3D in nature, as
opposed to the purely two-dimensional spin wave dispersion on the parent
cuprates. For the cuprates the spin fluctuations in the superconducting
regime are again purely 2D \cite{wilson,ncco,lake}, while the iron-based
superconductors appear to exhibit anisotropic 3D behavior like their parents.

The neutron scattering measurements were carried out on the SPINS cold and
BT-7 thermal triple-axis spectrometers at the NIST Center for Neutron
Research. We label the momentum transfer ${Q=}(q_{x},q_{y},q_{z})$ as $(H,K,L)=(q_{x}a/2\pi ,q_{y}b/2\pi ,q_{z}c/2\pi )$ reciprocal lattice units
(rlu) using the orthorhombic magnetic unit cell of the parent undoped
compound (space group $Fmmm$, $a=5.564$, $b=5.564$, and $c=12.77$~\AA ) for
easy comparison with previous spin wave measurements on the parent
compounds, even though the actual crystal structure is tetragonal \cite{jzhao3,mcqueeney,matan}. Many single crystals were co-aligned to obtain a
total mass of $\sim $1.2 grams. The in-plane and out-of-plane mosaics of the
aligned crystal assembly are $1.3^{\circ }$ and $4.3^{\circ }$ full width at
half maximum (FWHM), respectively \cite{chi}. For the experiment, the BaFe$_{1.9}$Ni$_{0.1}$As$_{2}$ crystal assembly was mounted in the $[H,0,L]$ zone
inside a liquid He cryostat. The final neutron wave vector was fixed at
either $E_{f}=5$ meV with a cold Be filter or at $E_{f}=14.7$ meV with a PG
filter in front of the analyzer. 

\begin{figure}[t]
\includegraphics[scale=.5]{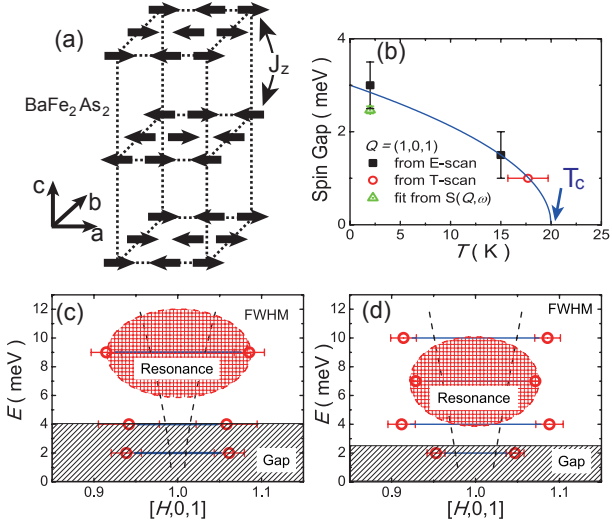}
\caption{(color online). (a) Schematic diagram of the Fe spin structure in
the BaFe$_{2}$As$_{2}$, which has magnetic Bragg peaks at $Q=(1,0,1)$, $(1,0,3)$, etc. For our experiment on BaFe$_{1.9}$Ni$_{0.1}$As$_{2}$, we use
the same unit cell for easy comparison. (b) Temperature dependence of the
spin gap as determined from energy scans (Fig. 3c) and the temperature
dependence of the scattering at $Q=(1,0,1)$ (Fig. 3d). The solid curve
represents the temperature dependence of the BCS gap function. (c,d)
Schematic of the magnetic response and spin gaps at $Q=(1,0,0)$, and $(1,0,1)
$, respectively. Measurements at $Q=(1,0,3)$ showed similar behavior as
those at $Q=(1,0,1)$.}
\end{figure}

\begin{figure}[t]
\includegraphics[scale=.5]{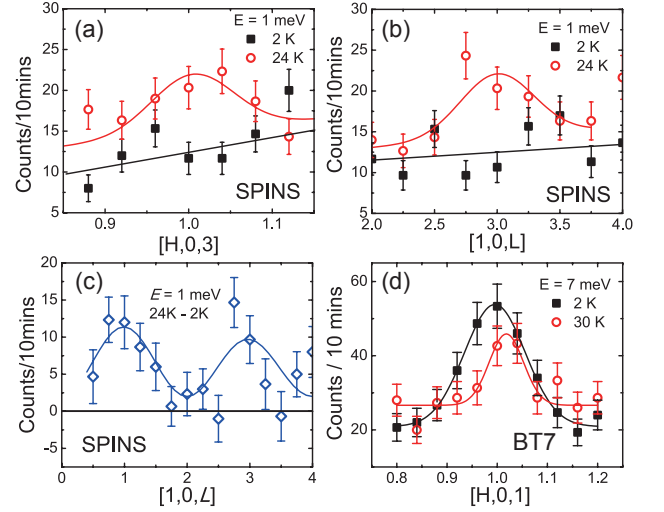}
\caption{(color online). Examples of constant energy scans around the $(1,0,3)$ position for $E=1$ meV obtained with $E_{f}=5$ meV above and below $
T_{c}$ on SPINS. (a) $Q$-scan along the $[H,0,3]$ direction for $E=1$ meV at
24 K and 2 K. A clear peak centered at $(1,0,3)$ at 24 K disappears at 2 K,
indicating the opening of a spin gap. (b) Similar scan along the $[1,0,H]$
direction showing a peak centered at $(1,0,3)$ that disappears below $T_{c}$. (c) Using scattering at 2 K as background scattering, we determine the
normal state $L$-modulation of the spin fluctuations by subtracting the 2 K
data from 24 K data. It is clear that spin fluctuations are 3D and have
similar modulations along the $c$-axis as spin waves. d) $Q$-scan in the
superconducting state through the magnetic resonance position, and above $T_{c}$ near $(1,0,1)$.}
\end{figure}

We first probe the wave vector dependence of the low-energy spin
fluctuations. Figures 2a and 2b show $[H,0,3]$ and $[1,0,L]$ scans at $E=1$
meV through the 3D $(1,0,3)$ Bragg peak position below and above $T_{c}$. We
see that the spin excitations observed above $T_{c}$ vanish at low $T$. \
Fig. 2c shows the intensity of the scattering above $T_{c}$ as a function of
wave vector along the \emph{c}-axis, using the low $T$ data as background,
and reveals the intrinsic wave vector modulation of the intensity of the
normal state spin fluctuations. The solid curve is a fit to the data using $\Delta S(Q,\omega )$(24 K$-$2 K)$=AF(Q)^{2}\sin ^{2}(\pi L/2)+C$, where $F(Q)
$ is the magnetic form factor of Fe$^{2+}$ and $C$ is constant. \ These data
are consistent with previous work on BaFe$_{1.9}$Ni$_{0.1}$As$_{2}$ which
showed that the spin fluctuation intensity has a $c$-axis modulation at $E=8.5$ meV, and a gap in the superconducting state \cite{chi}. \ For
comparison, Fig. 2d shows the magnetic scattering through the $[1,0,1]$
position in the superconducting state at the resonance energy of $E=7$ meV,
and the magnetic scattering above $T_{c}$. We note that in the undoped AF
state, the spin wave spectrum in BaFe$_{2}$As$_{2}$ has a gap of 9.8~meV 
\cite{matan}, while in the normal state of the doped system we find that the
spin fluctuation spectrum is gapless.

The behavior of the low energy spin excitations as a function of temperature
is shown in Fig. 3, which summarizes the BT-7 and SPINS data around $Q=(1,0,1)$. Figure 3a shows wave vector $[H,0,1]$ scans through the $Q=(1,0,1)$ position above and below $T_{c}$ at $E=2$ meV. A clear Gaussian
peak centered at $Q=(1,0,1)$ in the normal state vanishes below $T_{c}$,
demonstrating that the spin gap $\Delta _{spin}>2$ meV. Figure 3b plots the
signal and background scattering along the $[1,0,L]$ direction for $E=2$ meV
at 30 K, where we find that the normal state scattering also peaks at 3D AF
wave vector positions. To determine the spin gap value at $Q=(1,0,1)$, we
carried out temperature dependent measurements at 2 K, 15 K, and 30 K using
SPINS. We find a clear reduction in scattering (net negative values in the
subtraction) below 3 meV and 1.5 meV at 2 K and 15 K, respectively. These
results show that the maximum magnitude of the spin gap at the $Q=(1,0,1)$
wave vector is 3 meV, and the energy gap is temperature dependent.

\begin{figure}[t]
\includegraphics[scale=.5]{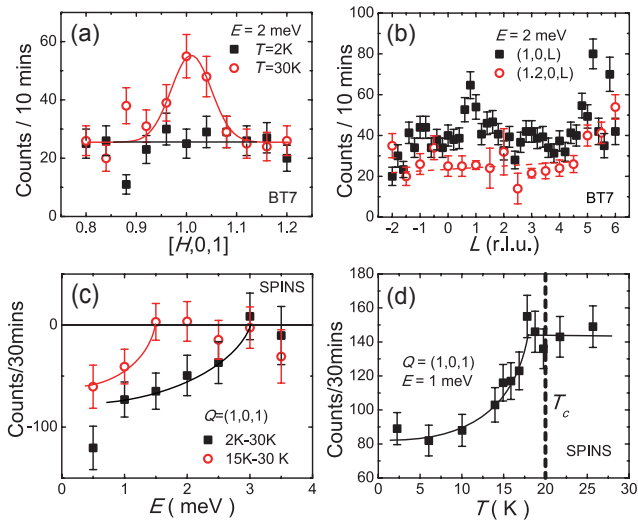}
\caption{(color online). (a) $Q$-scans at the $[H,0,1]$ direction above and
below $T_{c}$ at $\hbar \protect\omega =2$ meV. The data show the opening of
a spin gap at 2 meV below $T_{c}$. (b) $Q$-scans along the $[1,0,L]$
(signal) and $[1.2,0,L]$ (background) positions, showing the $L$ modulation
of the intensity, with maxima at $(1,0,1)$ and $(1,0,3)$. (c) Constant-$Q$
scans at $Q=(1,0,1)$ at various temperatures. The differences between low
and high temperature data show negative scattering due to the opening of a
spin gap. The data suggest a spin gap value of 1.5 meV at 15 K and 3.0 meV
at 2 K. (d) Temperature dependence of the scattering at $Q=(1,0,1)$ and $E=1$
meV shows a sudden drop below 18 K ($=T_{c}-2$ K) suggesting that the $E=1$
meV spin gap opens at a temperature slightly below $T_{c}$.}
\end{figure}

\begin{figure}[t]
\includegraphics[scale=.5]{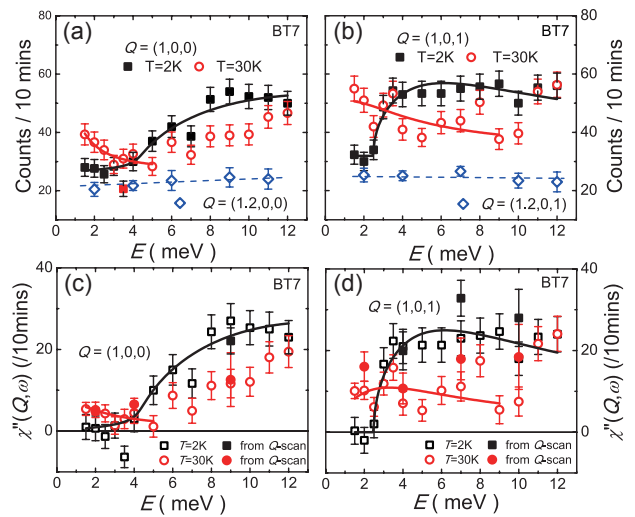}
\caption{(color online). Constant-$Q$ scans around the (a) $Q=(1,0,0)$ and
(b) $(1,0,1)$ positions above and below $T_{c}$, showing the developement of
the spin gap at low energies, and the enhancement of the magnetic scattering
at the resonance energy at each wave vector. Background data are indicated
by the open diamonds and dashed curves, and were collected at $Q=(1.2,0,0)$
and $Q=(1.2,0,1)$, respectively. (c, d) $\protect\chi ^{\prime \prime }(Q,\protect\omega )$ above and below $T_{c}$ obtained by subtracting background
and removing the thermal factor (see text). Also shown are values obtained
from the $Q$-scans at various energies above and below $T_{c}$. At both wave
vectors there is a clear magnetic intensity gain at the resonace energy of $
E=9.0$ meV at $Q=(1,0,0)$ and $7$ meV at $Q=(1,0,1),$ and a spin gap of 4.3
and 2.5 meV, respectively. The solid lines show fits using the model
described in the text.}
\end{figure}

To quantitatively determine the wave vector dependence of the spin gap in
the superconducting state, we carried out constant-$Q$ scans at the $
Q=(1,0,0)$, and $(1,0,1)$ wave vectors, and collected background data at $
Q=(1.2,0,0)$, $(1.2,0,1)$, above and below $T_{c}$ (Figs. 4a and 4b). In the
normal state (open circles) the magnetic scattering above background at both 
$Q=(1,0,0)$ and $(1,0,1)$ appears to increase with decreasing energy near
the elastic line, and thus suggests that this component of the scattering is
quasielastic in nature (peaks at $E=0$). In the superconducting state (solid
squares), the low energy scattering is suppressed, while the higher-energy
scattering increases in intensity. The overall behavior of the data is
remarkably similar to that in the optimally hole-doped La$_{1-x}$Sr$_{x}$CuO$
_{4}$ \cite{lake} and electron-doped Nd$_{1.85}$Ce$_{0.15}$CuO$_{4}$ \cite
{ncco}. However, it is also clear that the spin-gap occurs at a lower energy
at $Q=(1,0,1)$ than for $Q=(1,0,0)$, which is quite different than the
cuprates \cite{lake}. Figure 4c,d presents the data in the form of the
dynamic susceptibility $\chi ^{\prime \prime }(Q,\omega )$, which is related
to $S(Q,\omega )$ through the (removal of the) detailed balance factor; $\chi ^{\prime \prime }(Q,\omega )=(1-\exp (-\hbar \omega /k_{B}T))S(Q,\omega
)$. Recall that the thermal population factor increases with decreasing
temperature, and this function is divided into $S(Q,\omega )$ to obtain $\chi ^{\prime \prime }(Q,\omega )$ [with $\chi ^{\prime \prime }(Q,\omega
=0)=0$]. The filled circles are $\chi ^{\prime \prime }(Q,\omega )$ obtained
from $Q$-scans as a consistency check. Upon entering the superconducting
state, the spectral weight is rearranged, with the suppression of low energy
spin fluctuations and the appearance of the neutron spin resonance at
energies above the spin gap. The present data are consistent with the
reported spin resonance values of $E=9$ meV for $Q=(1,0,0)$ and $E=7$ meV
for $Q=(1,0,1)$ \cite{chi}. We estimate that the intensity of the resonance
is approximately compensated by the opening of the spin gap below the
resonance.

To quantify the magnitude of the spin gaps at $Q=(1,0,0)$ and $(1,0,1)$ in
the superconducting state, we follow previous work \cite{lake} and fit the
data with 
\begin{equation}
S(Q,\omega )={\frac{{AE^{\prime }\Gamma }}{{(\Gamma ^{2}+(\hbar \omega
)^{2})(1-\exp (-\hbar \omega /k_{B}T))}}},
\end{equation}%
where $E^{\prime }=Re{[(\hbar \omega -\Delta +i\Gamma _{s})(\hbar \omega
+\Delta +i\Gamma _{s})]^{1/2}}$, $A$ is the amplitude, $\Delta $ is the spin
gap, $\Gamma $ is the inverse lifetime of the spin fluctuations with $\hbar
\omega \gg \Delta $, $E^{\prime }$ is an odd function of $E=\hbar \omega $,
and $\Gamma _{s}$ is the inverse lifetime of the fluctuations at the gap
edge. \ The solid curves are the results of these fits. In the normal state,
this functional form does not provide an adequate fit over the entire energy
range, and we restricted it to lower energies (as indicated by the extent of
the curve for those data). We find $\Delta =0$ for both $Q=(1,0,0)$ and $(1,0,1)$. On cooling into the superconducting state, Eq. (1) can be used
over the entire energy range of the data, and the least-squares fit to the $Q=(1,0,0)$ data (solid curves in Figs. 4a and 4c) yields $A=56.7\pm 7.9$, $\Gamma =13\pm 6.5$ meV, $\Delta =4.3\pm 0.8$ meV, $\Gamma _{s}=0\pm 0.73$
meV. Similarly, for $Q=(1,0,1)$ we find $A=55.5\pm 14.5$, $\Gamma =5\pm 0.7$
meV, $\Delta =2.5\pm 0.08$ meV, $\Gamma _{s}=0\pm 0.53$ meV (solid curves in
Figs. 4b and 4d). The results of this analysis show that the the
superconducting spin gap values for $Q=(1,0,0)$ and $(1,0,1)$ are
distinctively different.

The present measurements, as well as the previous data on this material \cite{chi}, demonstrate that the resonance occurs at $E=9$ meV for $Q=(1,0,0)$,
which has a spin gap $\Delta =4.3\pm 0.8$ meV. \ For $Q=(1,0,1)$ the
resonance is at the lower energy of $E=7$ meV, and the spin gap also occurs
at the lower energy of $\Delta =2.5\pm 0.08$ meV. Therefore these two energy
scales track one another, with a ratio that is the same within the
uncertainties of the experiments. This is the expected behavior for the
singlet-triplet transition of a Cooper pair \cite{eschrig}.

We summarize in Figs. 1b-1d the key results of our experiments. The measured
temperature dependence of the spin gap at $Q=(1,0,1)$ is shown in Fig. 1b.
The solid curve shows the prediction of a simple BCS gap function near $T_c$, $\Delta
(T)=A(1-(T/T_{c}))^{1/2}$, which describes the data fairly well. Figures 1c
and 1d plot schematically the spin gap and resonance at $Q=(1,0,0)$ and $(1,0,1)$. The two energies exhibit the same dependence on wave vector. In
ARPES  experiments \cite{terashima}, two isotropic superconducting gaps with
values of 7 meV and 4.5 meV were observed for BaFe$_{1.85}$Co$_{0.15}$As$_{2}
$ with $T_{c}=25.5$ K. Comparison with the $Q=(1,0,0)$ neutron measurements
suggests that the resonance energy at $Q=(1,0,0)$ is indeed less than twice
the superconducting gap energy. These results are consistent with the idea
that the resonance is a bond state related to singlet-triplet excitations of Cooper
pairs, with a superconducting gap that varies with the momentum transfer
along the \emph{c}-axis \cite{eschrig}.

We thank Songxue Chi, Jun Zhao, and Leland Harriger for coaligning some of
the single crystals used in the present experiment. This work is supported
by the U.S. DOE BES No. DE-FG02-05ER46202, NSF DMR-0756568, and in part by the U.S. DOE,
Division of Scientific User Facilities. The work at Zhejiang University is
supported by the NSF of China.


\end{document}